\documentclass[%
 aip,
rsi,%
 amsmath,amssymb,
preprint,%
]{revtex4-1}

\usepackage{graphicx}
\usepackage{subcaption}
\usepackage{xcolor}
 \newcommand*{\citen}[1]{%
  \begingroup
    \romannumeral-`\x 
    \setcitestyle{numbers}%
    \cite{#1}%
  \endgroup   
}

\usepackage{dcolumn}
\usepackage{bm}

\def\er{\varepsilon_\textrm{r}}
\def\ur{\mu_\textrm{r}}
\def\an{a_\textrm{j}}
\def\bn{b_\textrm{j}}
\def\Qb{Q_\textrm{b}}
\def\Qsca{Q_\textrm{sca}}
\def\Rb{R_\textrm{b}}
\def\l0{\lambda}
\def\thetaeff{\theta_\textrm{eff}}

\usepackage{ulem}

\renewcommand{\sout}[1]{\unskip}

\begin{document}


\title{Experimental demonstration of spectrally-broadband Huygens sources using \sout{nonmagnetic} {{low-index}} spheres}

\author{Mohamed Ismail Abdelrahman}
 \altaffiliation[Current Address: ]{ School of Electrical and Computer Engineering, Cornell University, Ithaca, USA}
  \email{ mia37@cornell.edu}
  \affiliation{ 
Institute of Theoretical Solid State Physics, Karlsruhe Institute of Technology, Karlsruhe, Germany}%
 \affiliation{ 
Aix Marseille Univ, CNRS, Centrale Marseille, Institut Fresnel, Marseille, France}%

\author{Hassan Saleh}%
 \affiliation{ 
Aix Marseille Univ, CNRS, Centrale Marseille, Institut Fresnel, Marseille, France}%
\affiliation{ 
Centre Commun de Ressources en Microondes CCRM, Marseille, France
}%

\author{Ivan Fernandez-Corbaton}
\affiliation{%
Institute of Nanotechnology, Karlsruhe Institute of Technology, Karlsruhe, Germany}%

\author{Boris Gralak}
 \affiliation{ 
Aix Marseille Univ, CNRS, Centrale Marseille, Institut Fresnel, Marseille, France}%

\author{Jean-Michel Geffrin}
 \affiliation{ 
Aix Marseille Univ, CNRS, Centrale Marseille, Institut Fresnel, Marseille, France}%

\author{Carsten Rockstuhl}
  \affiliation{ 
Institute of Theoretical Solid State Physics, Karlsruhe Institute of Technology, Karlsruhe, Germany}%
\affiliation{%
Institute of Nanotechnology, Karlsruhe Institute of Technology, Karlsruhe, Germany}%





\begin{abstract}

Manipulating the excitation of resonant electric and magnetic multipole moments in structured dielectric media has unlocked many sophisticated electromagnetic functionalities. This invited article demonstrates the experimental realization of a broadband Huygens' source. This Huygens' source consists of a spherical particle that exhibits a well-defined forward-scattering pattern across more than an octave-spanning spectral band at GHz frequencies, {{where the scattering in the entire backward hemisphere is suppressed}}. Two different {{low-index nonmagnetic}} spheres are studied that differ in their permittivity. This causes them to offer a different shape for the forward-scattering pattern. The theoretical understanding of this broadband feature is based on the approximate equality of the {{resonant}} electric and magnetic multipole moments in both amplitude and phase in low permittivity spheres. This is a key condition to approximate the electromagnetic duality symmetry which, together with the spherical symmetry, suppresses the backscattering. With such a configuration, broadband Huygens' sources can be designed even if magnetic materials are unavailable. This article provides guidelines for designing broadband Huygens' sources using \sout{nonmagnetic} {{low-index}} spheres that could be valuable to a plethora of applications.

\end{abstract}

\maketitle

%

\section{Introduction} \label{sec:intro}

Engineering the light scattering at \sout{subwavelength} {{nanoscale}} dimensions has paved the way for a plethora of unprecedented functionalities such as nanoantennas,  lensing  using  planar surfaces, optical trapping, and advanced  optical imaging \cite{muehlschlegel2005resonant,pendry2000negative,arita2018invited,hu2018invited}. These applications come in reach due to the rich behavior of light scattering in the resonant regime, which we usually enter when the characteristic size of the scattering system is comparable to the wavelength of the excitation; that is  a vastly growing branch of nanophotonics  coined ``Meta-optics'' \cite{kivshar2017meta}. 
A large share of investigations targets to eliminate the {{total backward scattering}} (or reflection) and it has recently gained  increasing interest due to its importance in a variety of applications like solar cells \cite{spinelli2012broadband,nishijima2016anti},  light harvesting devices \cite{alaee2017theory}, and the unidirectional propagation in integrated optics and optical waveguides \cite{gangaraj2018topological}. 

\begin{figure}[htbp]
\centering
{\includegraphics[width=0.7\linewidth]{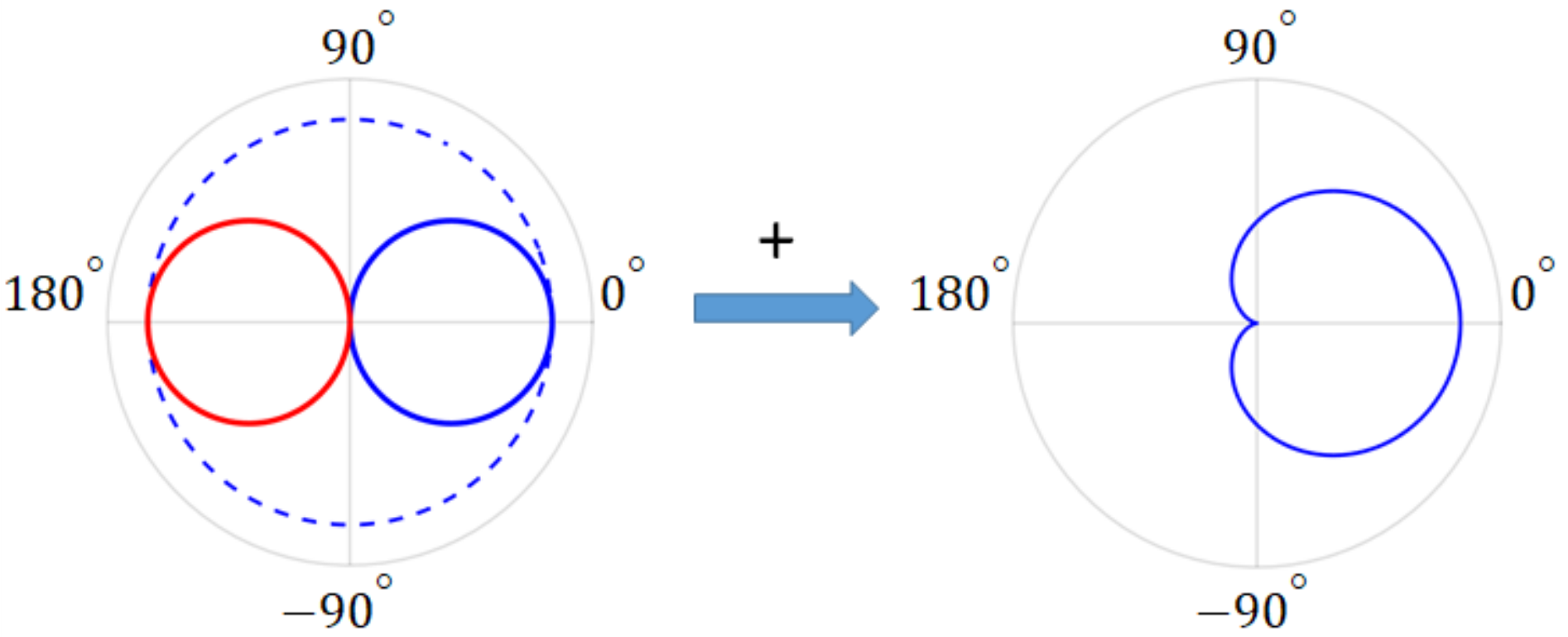}}
\caption{ Zero-backscattering  (in the backward direction: $180^{\circ}$)  due to the superposition between equally-excited electric/magnetic dipoles (dashed/line) of  even/odd parities of the amplitude distribution between the forward and backward direction. The polarization of the excitation is perpendicular to the shown  plane. }
\label{fig:ZBS}
\end{figure}

One of the initial attempts was to achieve zero-backscattering (ZBS) in the stealth technology \cite{wagner1963theorem}. There, it was first proposed that an impinging plane wave propagating parallel to the axis of symmetry of a rotational-symmetrical scatterer experiences no reflection  (invisible to radar) if the particle is made from an electromagnetic dual material. Indeed, a rotational symmetry, of at least the third order, is enough to suppress the backscattering for a dual scatterer \cite{fernandez2013electromagnetic}.   
Dual materials have equal electric permittivity and magnetic permeability $(\er=\ur)$; a condition widely known as the first Kerker condition for ZBS  \cite{kerker1983electromagnetic} originally proposed for magnetic spheres and later extended to other symmetrical shapes \cite{zambrana2013duality,fernandez2013forward}. In Mie theory \cite{mie1908beitrage}, ZBS is achieved through the destructive interference of every excited electric multipole moment, of amplitude coefficient $\an$, and its magnetic counterpart moment $\bn$ of the same multipole order $j$ but different parity of the scattering pattern \cite{bohren2008absorption}. See Fig.~\ref{fig:ZBS} as an example of the \textit{dipolar duality}  ($a_1=b_1$).  The condition $\an=\bn$ can be  satisfied for every $j$ only if $\er=\ur$ \cite{kerker1983electromagnetic}.  

The absence of magnetic materials at optical frequencies, and indeed in many other frequency ranges, however, renders it impossible to achieve ZBS using the Kerker condition as it was primarily proposed. Only recently,  pioneering works have  demonstrated a strong magnetic dipole resonance in resonant nonmagnetic silicon nanoparticles  at the visible range \cite{kuznetsov2012magnetic,person2013demonstration}. 
This artificial magnetism is due to the circular currents inside the nanoparticles rather than an intrinsic magnetic property. In this way, and by carefully tuning the particle size, ZBS can be achieved at a particular wavelength by equally exciting electric and magnetic dipole moments in both amplitude and phase \cite{geffrin2012magnetic,person2013demonstration}, assuming the contribution of higher order multipoles can be neglected.

These findings set the scene for further generalizations of the condition of ZBS to include higher-order multipoles \cite{alaee2015generalized,abdelrahman2017broadband} and, moreover, to investigate ZBS at different  configurations like nanodisks, core-shell particles, dimers, and dielectric metasurfaces \cite{staude2013tailoring,liu2012broadband,barreda2017electromagnetic,barreda2018scattering,decker2015high}. 
In all these cases, the underlying concept is always that the electric and magnetic polarizabilities are nearly the same so that the duality symmetry is preserved for the whole system (\textit{generalized Kerker condition}). Eventually, the interference of the excited electric and magnetic multipoles should be engineered so that the radiation vanishes in the backward direction at the desired wavelengths without the explicit need for magnetic materials, refer to Fig.~1 in the recent review paper [\citen{liu2018generalized}] for an elaborated illustration.

As previously mentioned, ZBS was demonstrated  at a single frequency that resembles the Kerker condition at the dipolar limit, yet this provides only a limited utility due to the narrow-band functionality \cite{geffrin2012magnetic,person2013demonstration}. Achieving ZBS across a broad spectrum is indispensable for many applications like antireflection coating  for solar cells and photodetectors. One possible approach to render the effect broadband is to investigate other particles with additional geometrical degrees of freedom compared to spheres, such as  disks, cubes, cones,  coated spheres, dimers \cite{evlyukhin2011multipole,rahimzadegan2017core,barreda2018scattering}. 
That allows achieving overlapped  electric and magnetic dipoles over a broad spectrum \cite{decker2015high}. Higher-order multipoles \cite{kruk2016invited} can also be involved to achieve broadband ZBS with a better directionality \cite{liu2014ultra}. 

 

Alternatively, it has been just recently explicitly pointed out that low-permittivity {{nonmagnetic}} spheres exhibit a large number of comparable {{resonant}} electric and magnetic multipole moments over a broad spectrum \cite{abdelrahman2017broadband}. They possess a nearly identical dispersion in both amplitude and phase. Therefore, instead of satisfying the rigorous Kerker condition of $\an=\bn$, it is possible to achieve $\an\approx\bn$ (\textit{approximate duality}) in an extended spectral region for a large number of multipole moments, i.e., up to the order $j=8$ in Ref.~[\citen{abdelrahman2017broadband}] with such spheres, even though nonmagnetic
materials are exclusively considered. That results in a broadband ZBS over an octave-spanning band of frequencies, and even more, accompanied by a large scattering cross-section {{(due to the resonant interaction with light)}} and a highly directive forward scattering, which is an unprecedented result for a single scatterer. Similar behavior has been reported for a cluster of colloidal particles \cite{dezert2017isotropic}.

These recent results suggest the possibility to {{go a step further to}} realize a spectrally-broadband Huygens' source (a forward-scattering particle) that suppresses the backscattering, {{not only in the backward direction but}} in the entire back hemisphere, based on the effect of the broadband approximate duality. In this article,  the first experimental verification at GHz frequencies is provided for a broadband Huygens' source using  \sout{sub} wavelength{{-sized}} nonmagnetic ($\ur=1$) spheres made of materials of permittivity $\er$ below $3$ ({{low-}}index $n = \sqrt{\er\,\ur} < 1.73$). The experimental  setup is similar to the setup  used to first verify Kerker conditions \cite{geffrin2012magnetic}. In addition, a parametric study is presented to provide guidelines for designing broadband Huygens' sources using spheres of different sizes and values of the permittivity.  {{Low-index materials have already shown its prominence in many optical components and applications \cite{luk2017refractive}}}.                

\section{Theoretical formulation}

The preliminary quantity to judge the backscattering, using Mie theory, is the backscattering efficiency\cite{bohren2008absorption} $\Qb$, which indicates the ability of a scattering system to scatter light in the direction opposite to the illumination ($\theta=180^\circ$). For spheres, $\Qb$ is expressed  in terms of the  Mie coefficients $\an$ and $\bn$ of the expansion on radiating spherical vector wave functions:
\begin{equation} \label{eq:Qb}
\Qb=  \frac{\lambda^2}{4 \pi^2 r^2}\; {\left| \sum_{j=1}^{\infty} (2j+1)\;(-1)^\textrm{j}\;(\an-\bn) \right|}^2,
\end{equation}
where $r$ is the sphere radius and $\lambda$ is the illumination wavelength in the surrounding medium. 
Mie coefficients  \cite{bohren2008absorption} are functions of the material properties of the sphere and the relative size of the sphere radius to the wavelength $r/\lambda$, which allows to easily scale the results to any desired spectral region. The approximate Kerker condition $\an\approx\bn$ requires $\er \approx$ $\ur$, which in turn leads to a broadband suppression of the backscattering  \cite{abdelrahman2017broadband}, i.e.,  $\Qb \approx 0$ and $\partial \Qb(\lambda)/ \partial\lambda\approx 0$. This necessary condition is met in good approximation for nonmagnetic spheres having a permittivity close to unity.


Investigating the backscattering alone could be misleading: another quantity must be taken into consideration, which is the strength of the scattering process itself. This is quantified by the scattering efficiency \cite{bohren2008absorption} $\Qsca$:
\begin{equation}
\Qsca =  \frac{\lambda^2}{2 \pi^2 r^2}\;  \sum_{j=1}^{\infty} (2j+1)\;( \left| \an \right|^2 + \left| \bn \right|^2 ).
\end{equation}

\begin{figure}[htbp]
    \centering
    \begin{subfigure}[b]{0.6\linewidth}
        \includegraphics[ width=1\linewidth, keepaspectratio]{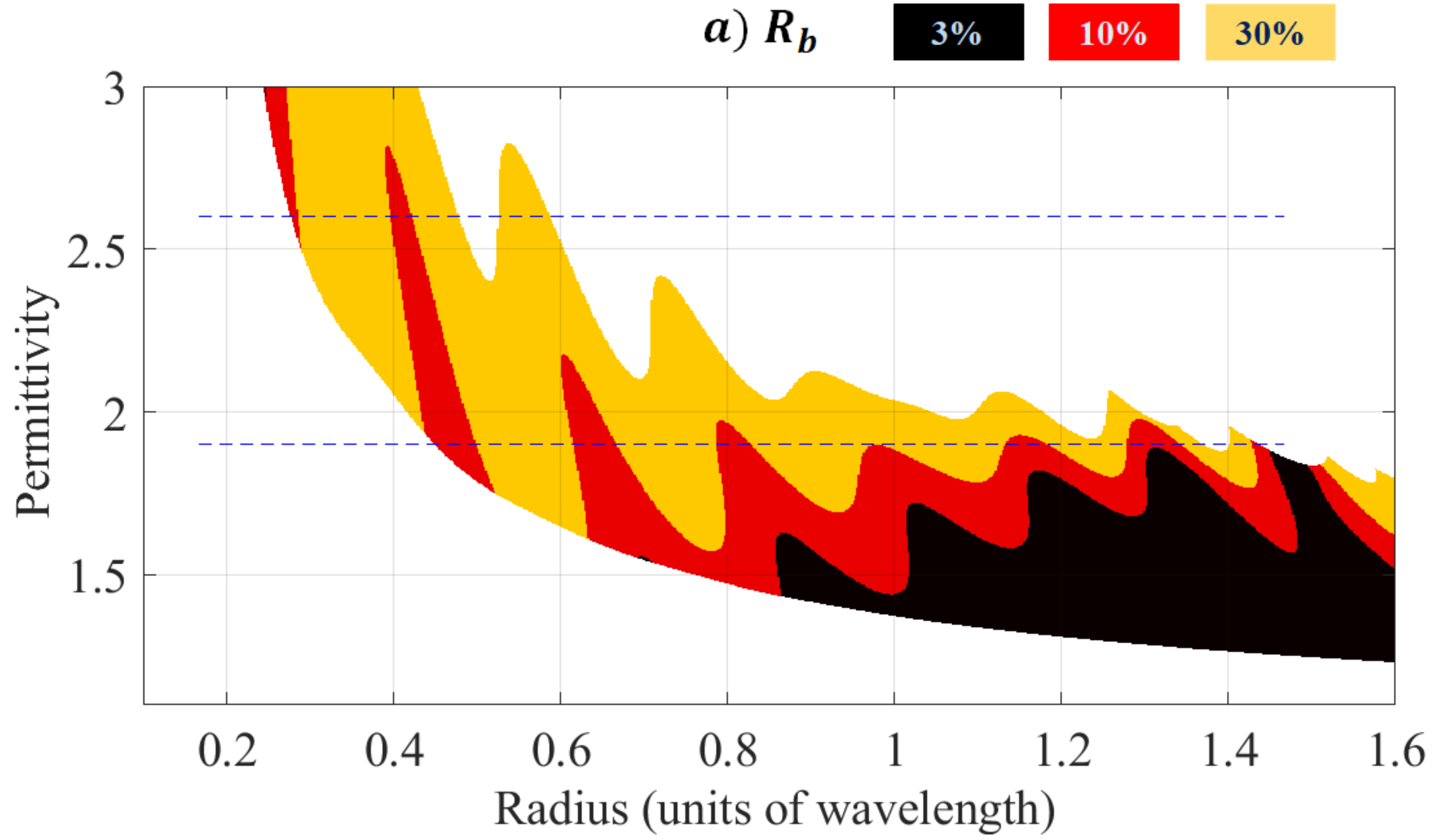}
    \end{subfigure} \\[2mm]
    \begin{subfigure}[b]{0.6\linewidth}
        \includegraphics[ width=1\linewidth, keepaspectratio]{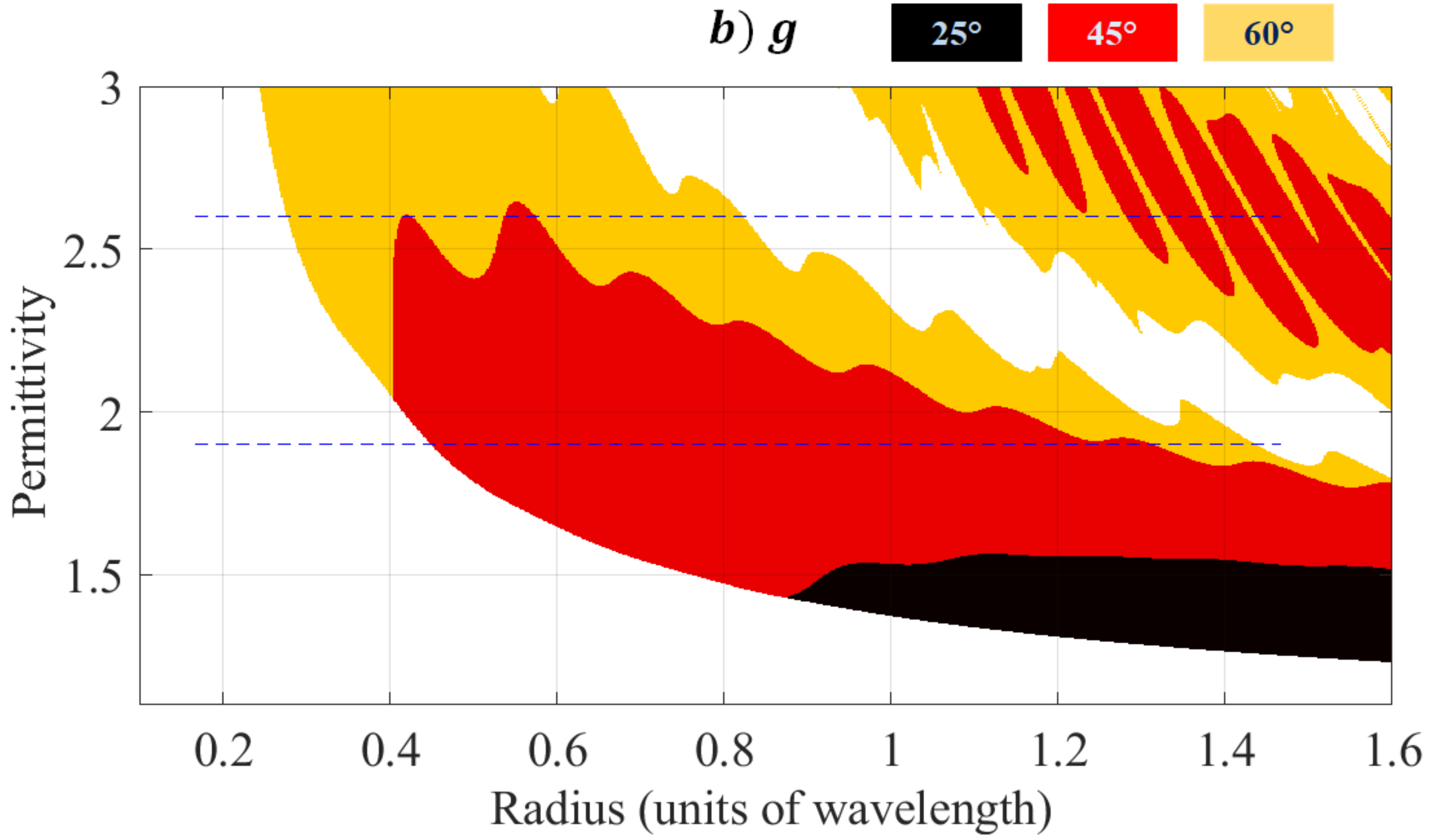}
    \end{subfigure}
\caption{Parameter regions  for a) Backscattering ratio $\Rb$ below $3\%,10\%,30\%$ and b) Asymmetry parameter $g \geq 0.9,0.7,0.5$ ($g=cos(\thetaeff): \thetaeff  \leq 25^\circ,45^\circ,60^\circ$), while the scattering efficiency  $\Qsca$ is  at least 2 {{(resonant interaction)}} in all the colored areas. The dashed lines represent the spheres under study of permittivity values $1.9$ and $2.6$. }
\label{Fig2:regions} 
\end{figure}

Broadband parameter regions can be therefore identified as to satisfy a low backscattering ratio defined here as $\Rb= \Qb / \Qsca$ given a non-negligible value of $\Qsca$, {{otherwise, the low backscattering is simply due to the absence of the interaction of light with the particle, which is a trivial case for low backscattering.}} Results of a parameter sweep are shown in Fig.~\ref{Fig2:regions}.a). The results highlight several potential regions of nonmagnetic subwavelength and wavelength-sized spheres 
that look suitable. The quantity $\Qsca$ is chosen to be at least 2 [3 dB], which characterizes the resonant scattering region \cite{bohren2008absorption}. The analysis is conducted by a freely available Mie scattering MATLAB code \cite{matzler2002matlab}, given that a sufficiently large number of multipole moments were retained to ensure convergence. The background material is always vacuum.

On the other hand, Huygens' sources are characterized by a  wide-angle forward scattering pattern, while the scattering in the whole backward hemisphere ($\theta:\,90^\circ\,\to 180^\circ\,\to\,-90^\circ$) nearly vanishes. 
The asymmetry parameter\cite{bohren2008absorption} $g$  is preferable to judge the backscattering in this case, being defined as the average (effective) cosine of the angular distribution of scattering function $X(\theta,\phi)$

\begin{equation}
g = \, \cos(\thetaeff)\, = \int_{4\pi} X(\theta,\phi) \, \cos(\theta)\, d\Omega,
\end{equation}
where $\theta$ is the angle in the scattering plane, $\phi$ is the azimuthal  angle, $\Omega$ is the solid angle, and    $X$ is a function of the scattering matrix elements $S_1(\theta,\phi)$ and $S_2(\theta,\phi)$ that can be evaluated directly for spheres using Mie coefficients, as given by section 4.5 in  Ref. [\citen{bohren2008absorption}].
 The function $g$ is independent of the polarization of the illumination in the case of  spherical particles and is ranged between $[-1,1]$, where $0$ indicates an isotropic scattering pattern, $1$ means only $\theta=0^{\circ}$ scattering and $-1$ only $\theta=180^{\circ}$.  Figure~\ref{Fig2:regions}.b) reveals several octave-spanning regions of wavelength-sized spheres of permittivity of  $1.4-2.1$ and subwavelength spheres of permittivity of  $1.8-2.9$, all exhibiting, effectively, a forward scattering pattern defined here as $g \geq 0.5$ ($\thetaeff \leq 60^\circ$), while $\Qsca \geq 2$ is ensured.

 It is important to note that  the parameter $g$ is not related, in principle, to the duality condition. However, and since we consider low  permittivity \sout{subwavelength} spheres where the scattering pattern only changes smoothly, ZBS is usually correlated with a reduced  scattering in the backward hemisphere \sout{for subwavelength spheres}. That is clearly evident by comparing the regions of ZBS in Fig.~\ref{Fig2:regions}.a) and the regions of Huygens' scattering  ($\thetaeff \leq 60^\circ$) in  Fig.~\ref{Fig2:regions}.b). 

The choice of the more suitable parameter $\Qb$ or $g$ to use depends on the application. Low backscattering efficiency $\Qb$ indicates a system of preserved duality symmetry which is an essential feature for realizing optical activity in an arbitrary  direction, optical sorting and many other applications \cite{fernandez2015dual,nieto2015optical,rahimzadegan2016optical,rahimzadegan2017core}. Alternatively, it is necessary to design a scattering coating with a wide-angle forward scattering, defined here as $ 0.7 \geq  g \geq 0.5$ ($45^\circ \leq \thetaeff \leq 60^\circ$) in order to increase the light trapping in solar cells \cite{atwater2010plasmonics}, while 
a needle-like  scattering in the forward direction of $g \geq 0.7$ ($\thetaeff \leq 45^\circ $) is required for tractor beams \cite{chen2011optical}, for instance.

\section{Experimental Results}
The experimental validation is done in the microwave domain. Insights can be directly translated to the optical domain by taking advantage of the scalability of Maxwell's equations. This requires the preservation of the wavelength/target size ratio as well as the electromagnetic properties of the target. 
 The measurements are carried out in the anechoic chamber of the Centre Commun de Ressources Micro-Ondes (CCRM). Microwave analog experiments have already been  allowed to make the experimental proof of Kerker conditions \cite{geffrin2012magnetic}.

\sout{For many years, the experimental facility of the CCRM has been used for carrying out microwave analog experiments on different kind of targets like high refractive index subwavelength spheres analogues of silicon particles to demonstrate the directional tunability of scattering radiations \cite{barreda2017electromagnetic}.
 Low refractive index spheroids analogs of microalgae in aquatic medium \cite{saleh2017microwave} and soot aggregates analogues of complex shape have also been investigated \cite{merchiers2010microwave} in CCRM. Through a scale reduction from UHF regime to microwave regime, it was also possible to make analog experiments of trees analogues representing a forest \cite{bellez2011rigorous}.}

The main features of the experimental facility are the possibility to control the position of the target under test as needed, the angles of incidence and reception, and the frequency. Therefore, with all the available mechanical displacement of the transmitting antenna and receiving antenna, it is possible to make measurements almost across the entire sphere surrounding the target. The experimental setup has already been described in several articles (see Refs.  [\citen{eyraud2009microwave},\citen{geffrin2009continuing}]). The target under test is placed at the center of the setup on a vertically expanded polystyrene support considered to be transparent to microwaves. A transmitting horn antenna is 1.7 meters away from the target and generates a linearly polarized  wave with a frequency ranges from 2 GHz to 18 GHz (or in the wavelength range $16-150$ mm).

\begin{figure}[htbp]
\centering
{\includegraphics[width=0.6\linewidth]{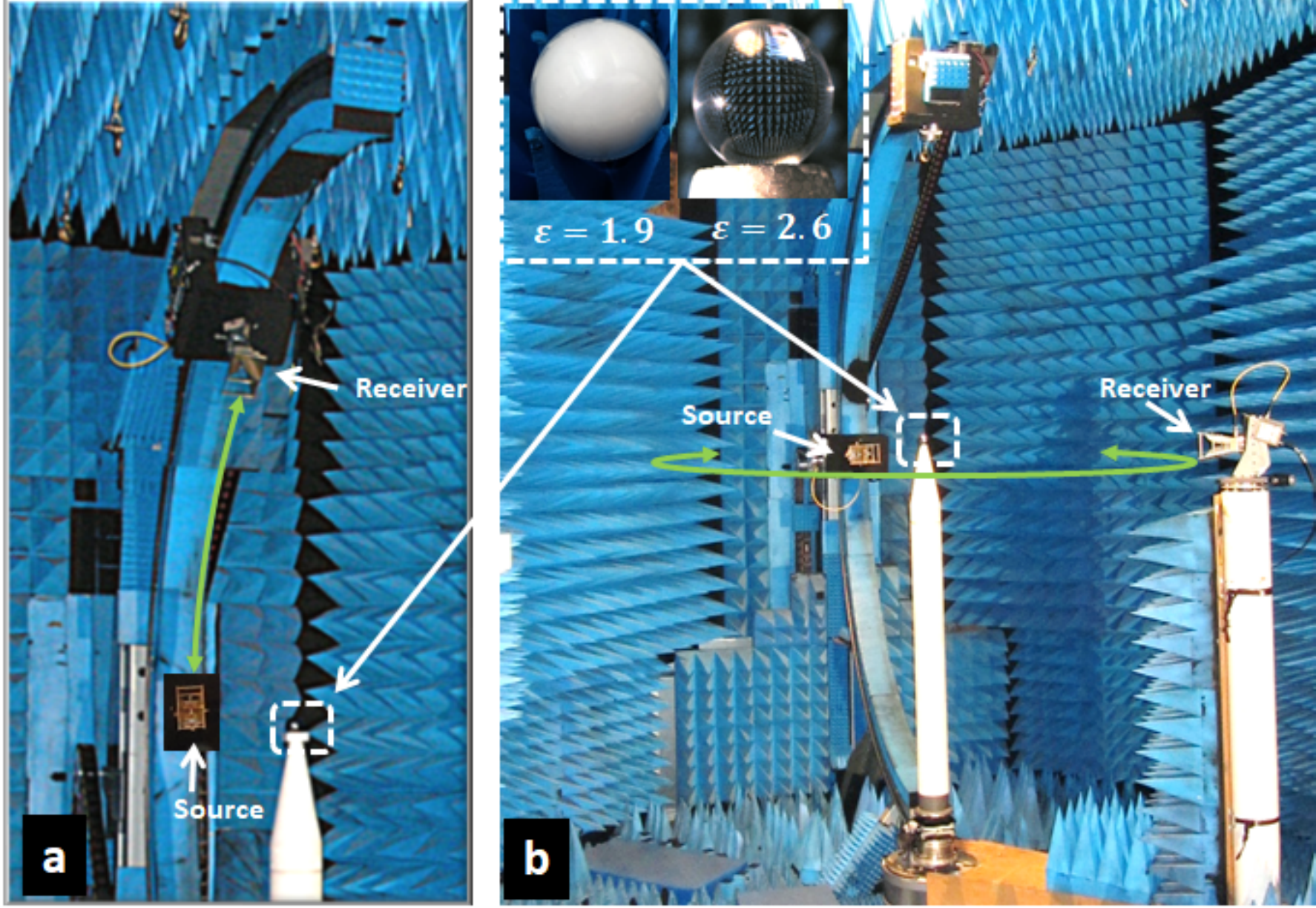}}
\caption{Photographs of the experimental microwave setup in the anechoic chamber showing the two configurations of measuring the full scattering pattern: a)  scattering angles between $-168^\circ$ and $-12^\circ$ and between $12^\circ$ and $168^\circ$ while the receiver is moving on a vertical arch, and b) scattering angles between $-130^\circ$ and $130^\circ$ while the receiver is moving horizontally. The target sphere is placed at the center of the setup, $1.7$ meters away from the source (at $180^\circ$). The insets show an actual photograph of the targeted spheres of permittivity $1.9$ ($n=1.38$) and $2.6$ ($n=1.61$).}
\label{fig:setup}
\end{figure}

Two geometrical configurations are used to measure the scattered field $E_\textrm{s}$. The first configuration allows the measurement for scattering angles  $-168^{\circ} \to -12^{\circ}$ (as well as the angles $12^{\circ} \to 168^{\circ}$) and it uses the displacement of the receiver on a vertical arch, as shown in Fig.~\ref{fig:setup}.a). The second configuration allows the measurement for scattering angles between $-130^{\circ}$ and $130^{\circ}$. It uses the displacement of the receiving antenna in the horizontal plane containing the source, the receiver and the target, see Fig.~\ref{fig:setup}.b). In both configurations, the receiver is 1.7 meters away from the target. Using both configurations allows constructing the full scattering pattern of the test object, except for a cone of $24^\circ$  around the backward direction since the receiver could physically block the transmitting signal. Nevertheless, the available scattering information is enough to fairly judge the directionality of the scattering pattern, especially that we are dealing with low refractive index spheres where the scattering patterns always change smoothly with the angle. More details about the two configurations can be found in Ref.~[\citen{geffrin2012magnetic}]. In both configurations, the polarization of the incident plane wave is perpendicular to the plane containing the source, the target and the receiver. 

The field scattered by the sphere cannot be obtained in a unique step but it is rather obtained from the complex subtraction of two different fields: the measured electric field with the sphere under test is at its place in the chamber (total field) and measured electric field in the absence of the sphere (incident field). After obtaining the scattered field, a drift-correction procedure is applied \cite{eyraud2006drift} to remove the drift errors that could happen due to the time delay between the total and incident field. In the end, the drift corrected scattered field is calibrated, using a measurement of a reference target, to refer the incident field to a magnitude equal to one and a null phase at the target's center and allow a quantitative comparison to simulations. 
    
\begin{figure}[htbp]
\centering
{\includegraphics[width=0.55\linewidth]{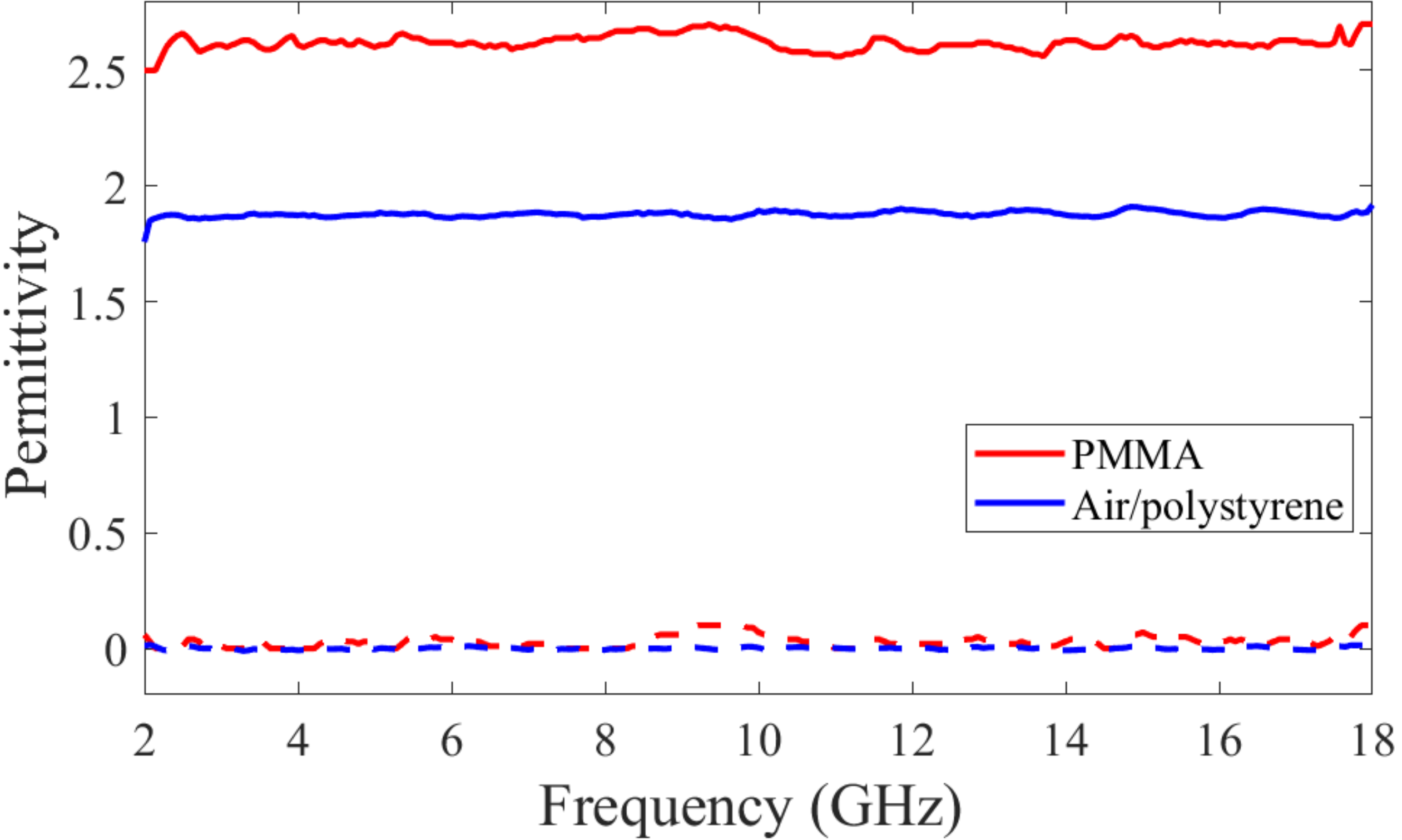}}
\caption{Characterization of the real (solid) and imaginary (dashed) parts of the  permittivity of {{both}}  spheres under study. {{The air/polystyrene sphere shows an effective real part of the permittivity around $1.9$, while the PMMA sphere permittivity is effectively around $2.6$. The characterization process shows that the imaginary part and the dispersion effect can be neglected across the whole frequency band under study.}} }
\label{fig:spheresccs}
\end{figure}

Two spheres have been chosen for the study, as marked by the dashed lines in Fig.~\ref{Fig2:regions}. The first sphere has $50$ mm diameter and is made of an air/polystyrene mixture of an effective real part of the permittivity around $1.9$. This sphere is expected to exhibit a narrow-band forward scattering, with a negligible backward scattering.  Alternatively, for a broadband Huygens' source (wide-angle forward scattering), the second sphere has $50.75$ mm diameter and made of PMMA material with and an effective real part of the permittivity of $2.6$.  To characterize the optical properties of each of the spheres, we rely on the iterative comparison between its measured scattered field and its calculated scattered field using Mie theory. The complex permittivity of the particle is chosen as the one corresponding to the best fit, in the least square sense, between Mie computations and the measured field, considering the measurement standard deviations. More details can be found in Ref.~[\citen{eyraud2015complex}]. Figure~\ref{fig:spheresccs} shows the characterized  permittivity of both spheres, and in both cases, the imaginary part and the dispersion effect can be neglected across the whole frequency band under study. \\

\begin{figure}[htbp]
    \centering
    \begin{subfigure}[b]{0.53\linewidth}
        \includegraphics[ width=1\linewidth, keepaspectratio]{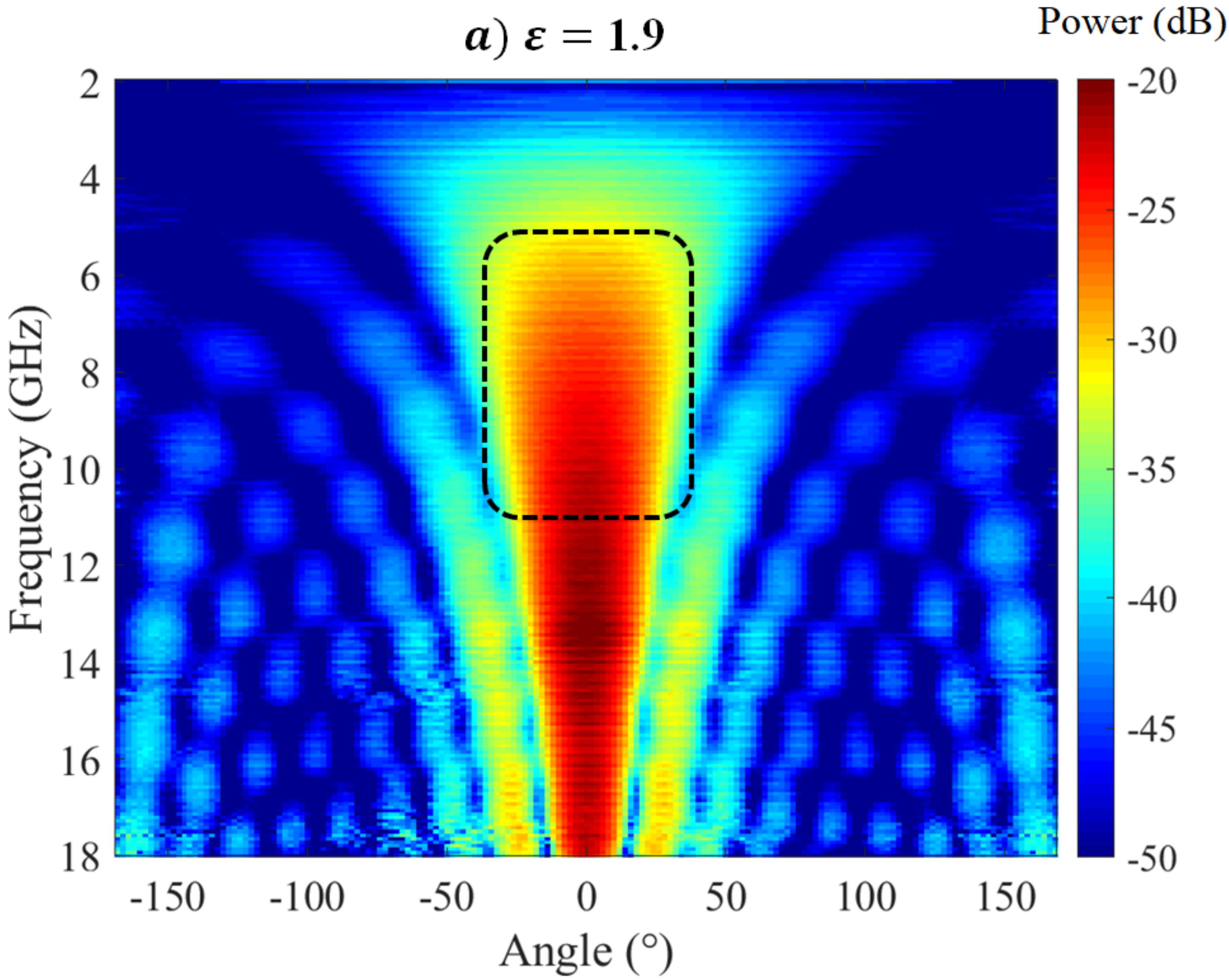}
    \end{subfigure} \\[2mm]
    \begin{subfigure}[b]{0.53\linewidth}
        \includegraphics[ width=1\linewidth, keepaspectratio]{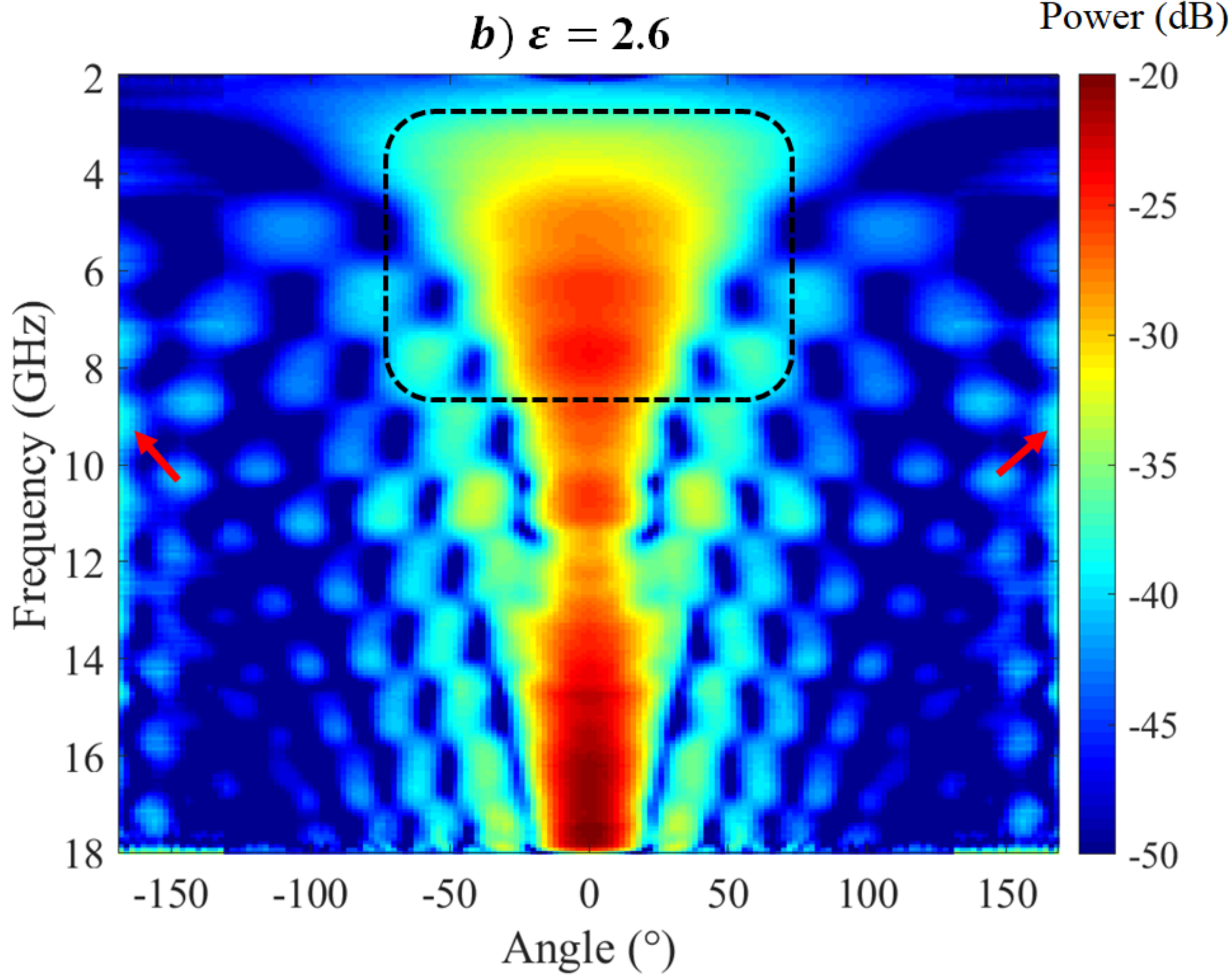}
    \end{subfigure}
\caption{Measurement results of the scattering pattern $20 \, \mathrm{log}_{10}(|E_\mathrm{s}(\theta)|)$ for $\theta$:$\{-168^{\circ},168^{\circ}\}$ across an extended spectrum $2$ GHz - $18$ GHz  for  a) a sphere of permittivity $1.9$ and diameter 50 mm and  b) a sphere of permittivity $2.6$ and diameter 50.75 mm. The dotted regions are examples of the parameter regions of interest that  exhibit a broadband a) narrow-angle and b) wide-angle forward scattering, accompanied by  $\Qsca \geq 2$. }
\label{Fig5:results} 
\end{figure}

The angular distribution of the power of the measured scattered fields (scattering pattern) of both spheres are shown in Fig.~\ref{Fig5:results}, for  a varying frequency. The results are shown in the dB scale that allows to identify both cases when the scattered power is distributed in a narrow or wide-angle, where in the linear scale, the latter case is overlooked due to the considerable disparity in the power per unit solid angle, assuming the total scattered power is the same in both cases.

\begin{figure}[htbp]
\centering
{\includegraphics[width=0.85\linewidth]{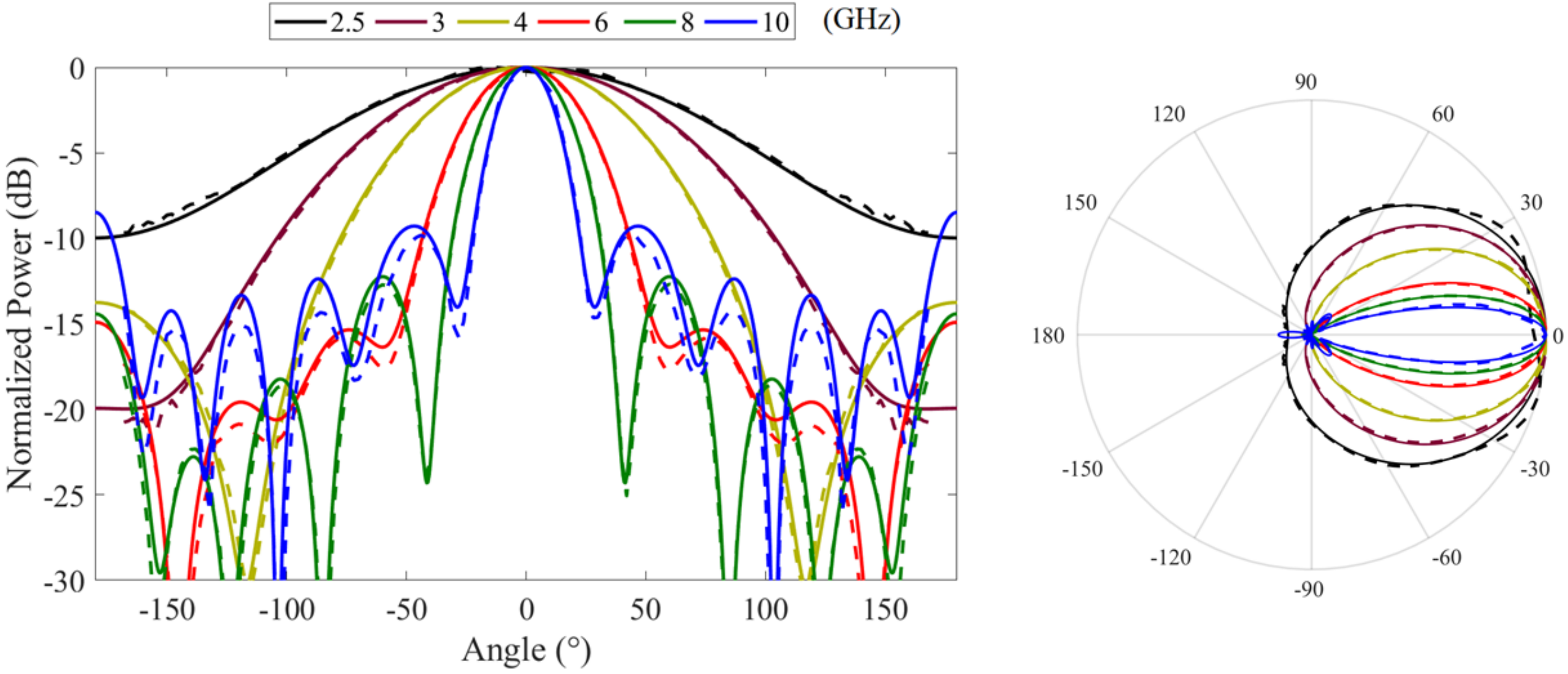}}
\caption{Scattering pattern (left: dB scale, right: polar coordinates-linear scale) at selected frequencies from Fig.~\ref{Fig5:results}.b) of the sphere of permittivity $2.6$ that exhibits  a Huygens' scattering pattern across a broad spectrum (legend shows the excitation frequency value in GHz). Results of Mie code (solid) show an excellent agreement with the measurement results (dashed). All the values of the scattering patterns are normalized to the scattering power at the forward direction $0^{\circ}$. }
\label{fig:ScatteringPattern}
\end{figure}

The dotted region in Fig.~\ref{Fig5:results}.a) for the sphere with the permittivity of $1.9$ is the region of interest that exhibits a  narrow-angle (single lobe) forward scattering, accompanied  by a high scattering efficiency as shown by the  calculations in Fig.~\ref{Fig2:regions}. As confirmed by the experiment, the broadband effect spans a broad spectral band (more than an octave band). Similarly, the dotted region in Fig.~\ref{Fig5:results}.b) for the sphere with the permittivity of $2.6$ is a spectrally broadband region of Huygens-like scattering pattern, as illustrated in Fig.~\ref{fig:ScatteringPattern}, where the wide-angle forward scattering is achieved for frequencies between $3-8$ GHz, with a vanishing scattering at the backward direction  as well. For higher frequencies,  non-negligible side lobes appears in the backward direction, as indicated by the arrows in Fig.~\ref{Fig5:results}.b). 

Figure~\ref{fig:ScatteringPattern} compares the normalized scattering pattern obtained by the measurement to the numerical results. Both show an excellent agreement, albeit for a little mismatching at the forward direction which can be explained by the glare effect; the high sensitivity of the scattered field extraction procedure at the forward direction that needs a subtraction of two large comparable quantities: the scattered fields  with and without the spheres. For a more elucidation, the scattering pattern is shown also in polar coordinates (linear scale).

\section{conclusion}

The first experimental demonstration of a spectrally-broadband Huygens's source using \sout{subwavelength} {{low-index}} nonmagnetic spherical particles \sout{made of low permittivity materials} is presented in this article. Parameter regions that achieve vanishing scattering in the backward hemisphere along with a high scattering efficiency {{(a signature of the resonant interaction with light)}} have been identified as a function of the materials permittivity and the sphere size, for different values of the effective scattering angle (or the asymmetry parameter). That allows  realizing different types of the angular distribution of the forward scattering pattern between the two limits of  the needle-like (narrow-angle) pattern and the Huygens-like (wide-angle) pattern. The theory behind this broadband effect is based on the  {{possibility of nonmagnetic spheres made of materials of low permittivity to support a large number of comparable resonant electric and magnetic multipole moments over a broad spectrum (approximate duality)  \cite{abdelrahman2017broadband}}}. This article provides a novel paradigm for the applications that require  subtle engineering of the scattering pattern over a broad spectrum of frequencies.

\section{Funding Information}
This work has been partly funded by the German Science Foundation within the priority program SPP1839 Tailored Disorder (RO 3640/7-2). The authors acknowledge the opportunity provided by the Centre Commun de Ressources en Microonde to use its fully equipped anechoic chamber.

\nocite{}
%

\end{document}